\documentclass[prd,preprintnumbers,twocolumn,superscriptaddress,nofootinbib]{revtex4-1}
\pdfoutput=1

\usepackage[utf8]{inputenc}
\usepackage[T1]{fontenc}
\usepackage{amsmath}
\usepackage{float}
\usepackage{graphicx}
\usepackage{comment}
\usepackage{xcolor}
\usepackage{color}
\usepackage{epsfig}
\usepackage{dcolumn}
\usepackage{hyperref}
\usepackage{bm}
\usepackage{amssymb}
\usepackage{slashed}
\usepackage{epstopdf}
\usepackage[normalem]{ulem}	
\pdfoutput=1

\makeatletter
\renewcommand{\maketag@@@}[1]{\hbox{\m@th\normalsize\normalfont#1}}%
\makeatother

\definecolor{red}{rgb}{1,0,0}
\definecolor{darkpink}{rgb}{0.8,0,0.4}

\DeclareRobustCommand{\Eq}[1]{Eq.~(\ref{#1})}
\DeclareRobustCommand{\Fig}[1]{Fig.~\ref{#1}}

\begin{document}

\title{N-Relaxion: Large Field Excursions from a Few Site Relaxion Model}

\author{N.~Fonseca}
\email{nayara@if.usp.br}
\affiliation{Instituto de F\'{i}sica, Universidade de S\~{a}o Paulo, SP, Brazil}
\affiliation{DESY, Notkestrasse 85, 22607 Hamburg, Germany}

\author{L.~de~Lima}
\email{leolima@ift.unesp.br}
\affiliation{DESY, Notkestrasse 85, 22607 Hamburg, Germany}
\affiliation{Instituto de F\'isica Te\'{o}rica, Universidade Estadual Paulista, SP, Brazil}

\author{C.~S.~Machado}
\email{camilasm@ift.unesp.br}
\affiliation{Instituto de F\'isica Te\'{o}rica, Universidade Estadual Paulista, SP, Brazil}
\affiliation{Theoretical Physics Department, CERN, Geneva, Switzerland}

\author{R.~D.~Matheus}
\email{matheus@ift.unesp.br}
\affiliation{Instituto de F\'isica Te\'{o}rica, Universidade Estadual Paulista, SP, Brazil}

\date{\today}

\begin{abstract}
 Relaxion models are an interesting new avenue to explain the radiative stability of the Standard Model scalar sector.  They require very large field excursions, which are difficult to generate in a  consistent UV completion and to reconcile with the compact field space of the relaxion. We propose an $N$-site model which naturally generates the large decay constant needed to address these issues. Our model offers distinct advantages with respect to previous proposals: the construction involves non-abelian fields, allowing for controlled high energy behaviour and more model building possibilities, both in particle physics and inflationary models, and also admits a continuum limit when the number of sites is large, which may be interpreted as a warped extra dimension.
\end{abstract}

\preprint{CERN-TH-2016-016 }
\preprint{DESY 16-011 }
\maketitle
%\tableofcontents

%%%%%%%%%%%%%%%%%%%%%%%%%%%%%%%%%%%%%%%%%%%%%%%%%%%%%%%%%%%
%%%%%%%%%%%%%%%%%%%%
\section{Introduction}
Large field excursions are known to be an ingredient of slow roll theories of inflation~\cite{*[][{ and references therein}] Lyth:2007qh,Baumann:2009ds}, and have become a requirement for relaxation solutions to the hierarchy problem of the Standard Model (SM)\cite{Graham:2015cka}. In these scenarios we have a scalar field starting at some large value  and slowly decreasing during the inflationary epoch. As an illustration, consider the relaxion model~\cite{Graham:2015cka, Espinosa:2015eda}:
{\small\begin{equation} \label{e:Vrelaxion}
V(\phi,H)=\Lambda^3 g\phi- \frac{1}{2}\Lambda^2 \!\left(1\!-\!\frac{g\phi}{\Lambda}\right)H^2+ \Lambda_c^4(H) \cos(\phi/f) +\cdots,
\end{equation}
}where $H$ is the Higgs field, $\Lambda$ is the cutoff of the model, $\phi$ is the relaxion field (assumed to be a pseudo-Nambu-Goldstone-Boson (pNGB) with decay constant $f$), the  spurion $g$ quantifies the explicit breaking of the discrete shift symmetry and $\Lambda_c(H)$ is a scale depending on the Higgs vev so that $\Lambda_c(H) \neq 0 \leftrightarrow \langle H \rangle \neq 0$.

It is technically natural to set $g$ to small values, so the first term in \Eq{e:Vrelaxion} is responsible for the slow roll of $\phi$. Once the coefficient of $H^2$ on the second term  becomes negative
 $H$  acquires a vev and one can show that the Higgs mass is much smaller than $\Lambda$.
 As $\Lambda_c(H) \neq 0$,
  $\phi$ gets trapped close to this phase transition (which fixes $\langle H\rangle$). If this is to work in a natural way
   we must assume $\phi$ scanned the typical range of field values $\Delta\phi \sim \Lambda / g \gg \Lambda$.

There are relevant concerns regarding this idea:% that we summarize below:

\begin{itemize}
\item
While having field excursions larger than the cutoff of the effective theory is not a problem in itself,
it might be problematic to construct a theory that could consistently
generate these large excursions, specially if the UV theory includes quantum gravity~\cite{ArkaniHamed:2006dz, Cheung:2014vva, Heidenreich:2015wga, Heidenreich:2015nta}.

\item
Another crucial feature of \Eq{e:Vrelaxion} is the presence of a linear term  that explicit breaks a gauge symmetry (the axion shift symmetry), which is inconsistent with the pNGB nature of the relaxion~\cite{Gupta:2015uea}.
\end{itemize}

This second point can be avoided if all operators involving $\phi$ are periodic, but with very different periods, and the linear term is nothing but a small region in an oscillation of longer period.  A simple way to  generate such oscillations is to produce a large hierarchy between the decay constants  \cite{Kaplan:2015fuy, Choi:2015fiu, Choi:2015aem, SlidesEspinosa:2015, Kim:2004rp, Harigaya:2014eta, Choi:2014rja, Higaki:2014pja, Harigaya:2014rga, Peloso:2015dsa}:
\begin{equation}\label{e:relaxion}
V(\phi, H) \sim  \Lambda^4 \cos{\left(\frac{\phi}{F}\right)} + \Lambda^4_c(H) \cos{\left(\frac{\phi}{f}\right)},
\end{equation}
where $F\gg f$. If additionally $F > \Lambda$ then  the first point is also addressed, because $\phi$ will have a compact field space of size $2 \pi F$ (we will comment on gravity related problems below).

An explicit example is proposed in~\cite{Kaplan:2015fuy} to generate  an effective super-Planckian field range, by considering $N+1$ complex scalars   with   the same decay constant  $f<M_{\rm{Pl}}$. By adding  a  conveniently chosen   breaking term, the global  $U(1)^{N+1}$ is  explicitly broken to $U(1)$ and the remaining pNGB has a decay constant which  exponentially depends on the number of fields as $F\gg e^{c N} f$, where $c \sim \mathcal{O}(1)$. It is emphasized in \cite{Kaplan:2015fuy}   that this construction cannot be interpreted as a deconstructed extra dimension, i.e. there is no continuum limit for this model. Other approaches achieving similar results are employed in~\cite{Ibanez:2015fcv, Hebecker:2015zss, Kappl:2015esy}.

In the following we present a different approach that can deal with the  issues discussed previously and at the same time indicates a different strategy to  search for  UV completions for the relaxation mechanism. The two main advantages of our approach are that: (i) the model does have a continuum limit that could be interpreted as an extra dimension; and (ii) we show that the desired features can be obtained from non-abelian groups, allowing for controlled (asymptotically free) UV behaviour.

A  concern arising when gravity is included in the UV theory is the  Weak Gravity Conjecture (WGC)~\cite{ArkaniHamed:2006dz}, which limits how small the coupling constants in gauge theories may be.
In a non-abelian setup, the conjecture is not yet sufficiently explored, however %, since embedded abelian Yang-Mills black hole solutions exist~\cite{Galtsov:1989ip, Volkov:1998cc, Bizon:1992pi, Kleihaus:1997rb}
it is expected that the usual arguments will also apply to this case~\cite{Galtsov:1989ip, Volkov:1998cc, Bizon:1992pi, Kleihaus:1997rb} . We leave this matter for future work.

Finally, it is important to see if one can find a viable and natural inflation model compatible with the relaxion scenario.
For explorations along these lines, see \cite{DiChiara:2015euo,Patil:2015oxa}.

%%%%%%%%%%%%%%%%%%%%%%%%%%%%%%%%%%%%%%%%%%%%%%%%%%%%%%%%%%%
%%%%%%%%%%%%%%%%%%%%
\section{\label{s:model} Minimal Model}
\begin{figure}[!h]
\includegraphics[width=\columnwidth]{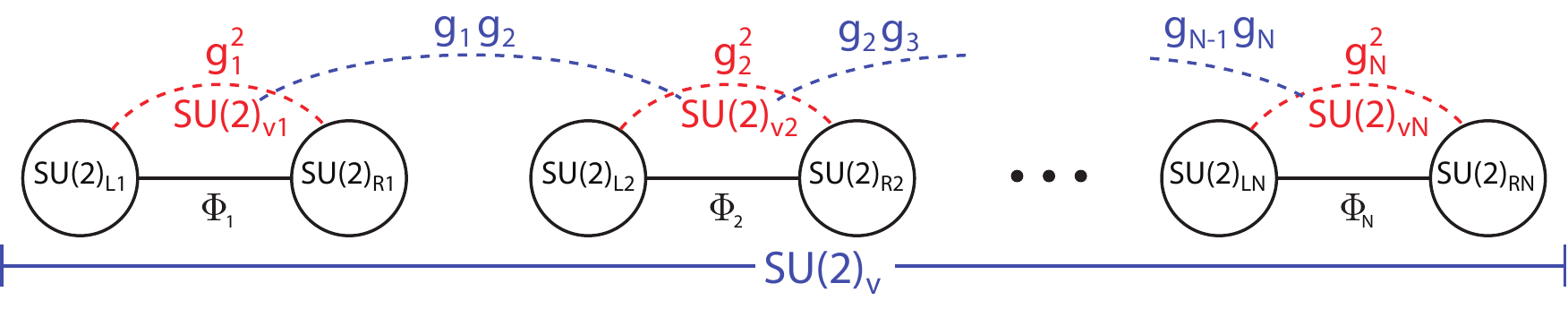}
\caption{\label{fig:bosons}
Diagram for a $2N$-site model. The symmetry groups and link fields are in black. In red (blue) we show the effect of the $g_j^2$ ($g_j g_{j+1}$) explicit breakings, and the resulting preserved groups.}
\end{figure}

We consider a $2N$-site model represented in  \Fig{fig:bosons}, where each site represents a global symmetry group \footnote{It is well known that in a theory of quantum gravity, all global symmetries are violated (see  e.g. \cite{Banks:2006mm}). For this reason, the model we propose in \Eq{e:Lphi} cannot be regarded as a consistent description for arbitrary energy scales. However, it may be seen as an effective few site description of an extra dimension (see Appendix \ref{app:pngb}). In this case, the global symmetries are gauged, and this concern disappears.}, $SU(2)$ (the construction is trivially generalized for other groups).

The Lagrangian for the link fields reads:
\begin{align}\label{e:Lphi}
\mathcal{L}_{\Phi}&=\!\sum_{j=1}^N \mathrm{Tr}\! \left[ \partial_\mu \Phi_j^\dag \,\partial^\mu \Phi_j\!+%\right.\nonumber\\
%&\left.+
\!\frac{f^3}{2}(2\!-\!\delta_{j,1}\!-\!\delta_{j,N})g_j^2 \!\left(\Phi_j\!+\!\Phi_j^\dag\right)\! \right]
\nonumber\\ &- \frac{f^2}{2}\sum_{j=1}^{N-1} g_j g_{j+1} \mathrm{Tr}\left[(\Phi_j- \Phi_{j}^{\dagger})(\Phi_{j+1}- \Phi_{j+1}^{\dagger} ) \right],
\end{align}
where the $\Phi_j$ are scalars transforming as  $\Phi_j \rightarrow L_{j}\Phi_j R^\dag_{j}$, under adjacent $SU(2)$ groups. We assume the $\Phi_j$ acquire a vev $\langle \Phi_j \rangle \equiv f/2$, spontaneously breaking $SU(2)_{L_j} \times SU(2)_{R_j}\rightarrow SU(2)_{V_j}$. In the low energy limit, these fields are non-linearly realized as:
\begin{equation}
\Phi_j \rightarrow
\frac{f}{2} e^{i \vec{\pi}_j \cdot \vec{\sigma}/ f}  = \frac{f}{2} \cos\! \left(\frac{\pi_j}{f}\right) + i \frac{f}{2} \frac{\vec{\pi}_j \cdot \vec{\sigma} }{\pi_j}\sin\!\left(\frac{\pi_j}{f}\right)\!,
\end{equation}
where $\vec{\sigma}$ are the Pauli matrices, $\vec{\pi}_j$ are the NGB multiplets and $\pi_j \equiv \sqrt{\vec{\pi}_j \cdot \vec{\pi}_j}$.

The Lagrangian contains terms that explicitly break some global symmetries. These parameters are assumed to be small spurions generated at a higher scale and may be chosen such that they give a mass to all but one linear combination of the $\vec{\pi}_j$. The terms with $g_j$ explicitly break the chiral symmetries to the vector combination, $SU(2)_{L_j}\times SU(2)_{R_j}\rightarrow SU(2)_{V_j}$, while the terms with $g_j g_{j+1}$ break $SU(2)_{V_j} \times SU(2)_{V_{j+1}}\rightarrow SU(2)_{V_{j,j+1}} $. Taken together these terms break explicitly all symmetries down to a diagonal $SU(2)_V$. However, due to the peculiar structure of the breaking parameters, one combination of the $\vec{\pi}_j$ remains accidentally lighter, gaining a small mass only at higher order. Additional breaking terms (involving three or more powers of the $\Phi_j$ fields) could be present, but we will assume that they are suppressed in relation to those in  \Eq{e:Lphi} (see Appendix \ref{app:UV} for an example of possible UV scenario).

The Lagrangian in terms of the Goldstone fields is:
\begin{align} \nonumber
\mathcal{L}_{\pi}&=\! \sum_{j=1}^N \left[ \frac{1}{2}\partial_\mu \vec{\pi}_j\!\cdot \!\partial^\mu \vec{\pi}_j \!+ %\right.\\ \nonumber
%&\left.
 \!f^4 (2\!-\delta_{j,1}\!-\!\delta_{j,N})g_j^2 \cos\! \left(\frac{\pi_j}{f}\right) %+\cdots
 \right]  \\ & + f^4 \sum_{j=1}^{N-1} g_{j}g_{j+1} \frac{\vec{\pi}_j \cdot \vec{\pi}_{j+1} }{\pi_j \pi_{j+1}} \sin \left( \frac{\pi_j}{f}\right) \sin\left(\frac{\pi_{j+1}}{f}\right),
 \end{align}
where we omitted terms corresponding to interactions with two derivatives. Expanding to quadratic order, we obtain the mass matrix for the $\vec{\pi}_j$, which is independent of the $SU(2)$ index:

 \begin{equation}
  \vec{\pi}^{T}\cdot M_\pi^2 \cdot \vec{\pi}\equiv
  \sum_{j=1}^{N-1} f^2(g_j \vec{\pi}_j-g_{j+1}\vec{\pi}_{j+1})^2,
 \end{equation}
 where $\vec{\pi}^T \equiv \{\vec{\pi}_1,\cdots,\vec{\pi}_N\}$.

The parametrization  $g_j \rightarrow q^j $, with $0<q<1$, results in a mass matrix for the pNGBs that is identical to the one obtained for a pNGB Wilson line (zero mode) in the deconstruction of $\text{AdS}_5$~\cite{tese:2013,Burdman:2014ixa} (see Appendix \ref{app:pngb})

\scriptsize
 \begin{equation} \label{m:MNG2}
M_\pi^2 = f^2 \left( \begin{array}{cccccc}
~~q^2     & -q^3     & ~~0          &\ldots   &~~0                 & ~~0 \\
-q^3      & 2q^4   & -q^5       &\ldots   &~~0                 & ~~0 \\
~~0       & -q^5     & 2q^6   &\ldots   &~~0                 & ~~0 \\
\vdots    & \vdots    & \vdots   &\ldots   &\vdots              & \vdots \\
~~0       & ~~0       & ~~0         &\ldots   &2q^{2(N-1)} & -q^{2N-1}   \\
~~0       & ~~0       & ~~0        &\ldots   &-q^{2N-1}     &q^{2N}
\end{array} \right).
\end{equation}
\normalsize
Note that since $\mathrm{Det}[M_\pi^2]=0$, this matrix has a zero mode (at tree level), as advertised. Its profile is given by:
\begin{equation}
\vec{\eta}_0= \sum_{j=1}^{N} \frac{q^{N-j}}{\sqrt{\sum_{k=1}^{N} q^{2(k-1)}}}\vec{\pi}_j,
\label{e:zeromode}
\end{equation}
which is similar to the result found in \cite{Kaplan:2015fuy}. One sees that $\vec{\eta}_0$ is exponentially localized at the last site. It is important to note that, in contrast with \cite{Kaplan:2015fuy}, since $q<1$ our matrix does admit a continuum limit, which should correspond to some bulk scalar in $\text{AdS}_5$.

Since $\vec{\eta}_0$ has a mass much smaller than the other states\footnote{At tree level, for $q \ll 1$, the spectrum is approximately given  by $m_j^2 \approx f^2 q^{2(j-1)}$ for $1 < j \leq N$ plus a zero mode. Expanding
 \Eq{e:Leta}, a quartic term is generated of order $q^{2 N}\! (\vec{\eta}_0\cdot\vec{\eta}_0)^2$. Closing the loop, one obtains a mass for $\vec{\eta}_0$ of order $m_{\eta_0}=f^2 q^{2 N}$, which is a factor of $q^2$ smaller than the lightest tree level mass, hence the approximation scheme is consistent.}, one is justified to consider it as the relaxion field, since the other modes rapidly lose coherence on scales larger than their Compton wavelength and may thus be assumed to be constant on the scale $m_{\eta_0}^{-1}$. They correspond to immaterial phase shifts in the potential of $\vec{\eta}_0$. In terms of $\vec{\eta}_0$, one obtains the following Lagrangian after integrating out the other pNGBs:
\begin{align} \label{e:Leta}
\mathcal{L}_{\eta}&=\sum_{j=1}^N  \left[ \frac{1}{2}\partial_\mu \vec{\eta}_0\cdot \partial^\mu \vec{\eta}_0 +f^4(2-\delta_{j,1}-\delta_{j,N}) q^{2j} \cos\frac{\eta_0}{f_j}\right]\nonumber \\
&+\sum_{j=1}^{N-1} f^4 q^{2 j+1}  \sin{\frac{\eta_0 }{f_{j}}} \sin{\frac{\eta_0}{f_{j+1}}},
 \end{align}
where $\eta_0 \equiv \sqrt{\vec{\eta}_0 \cdot \vec{\eta}_0}$ and  the effective decay constants are given by:
\begin{equation}\label{e:fjota}
  f_j \equiv  f \frac{\sqrt{\sum_{k=1}^N q^{2(k-1)}}}{q^{N-j}} \equiv f q^{j-N} \mathcal{C}_N,
\end{equation}
where $\mathcal{C}_N \equiv \sqrt{q^{2N} - 1 \over q^2 -1}$. One sees that a large hierarchy of decay constants is generated, from the largest $f_{\text{max}}=f_1 \approx f/q^{N-1}$ to the smallest $f_{\text{min}}=f_N \approx f$, as we wanted.

Regarding the radiative stability of the potential, we find that interactions  with $m$ external $\vec{\eta}_0$ legs scale as $c_m \sim q^{2 N} f^{4-m}$ and renormalize multiplicatively (as expected, since all the couplings in the Lagrangian \Eq{e:Leta} are spurions), so the whole potential is radiatively stable up to small corrections.

%%%%%%%%%%%%%%%%%%%%%%%%%%%%%%%%%%%%%%%%%%%%%%%%%%%%%%%%%%%
%%%%%%%%%%%%%%%%%%%%
\section{\label{s:relaxion} Higgs-Axion Interplay}

If the lightest pNGB  is to function as a relaxion, its potential must be such that no local minima stops it when the Higgs vev is zero.
The potential in \Eq{e:Leta} is dominated by the oscillation with the largest amplitude and period, $-f^4 q^{2} \cos{\eta_0 \over f_1}$, which grows monotonically in $0<\eta_0<\pi f_1$ (which will be our region of interest). To check that the other oscillations do not get the field stuck we need to consider:
\begin{align}\label{e:derV}
\frac{\partial V_{\eta}}{\partial \eta_0} = {f^3 q^N \over \mathcal{C}_N}
\sum_{j=1}^{N} q^j \sin{\left(\eta_0\over f_j\right)}
\bigg\{
    (2-\delta_{j,1}-\delta_{j,N})\,+
    \nonumber\\
    - (1-\delta_{j,1}) \cos{\left(\eta_0\over f_{j-1}\right)}\! - \!(1-\delta_{j,N}) \cos{\left(\eta_0\over f_{j+1}\!\right)}
\bigg\}.
\end{align}

The constant $f^3 q^N \over \mathcal{C}_N$ is positive for any $q < 1$ and $N > 1$, and the term between braces is bounded between $0$ and $4$ ($0$ and $2$ for $j=1$ and $j=N$). The leading term for small $q$ is:
\begin{equation}\label{e:derV1}
{f^3 q^N \over \mathcal{C}_N}
 q\sin{\left(\eta_0\over f_1\right)}
 \bigg\{
    1
    - \cos{\left(\eta_0\over f_{2}\right)}
 \bigg\}
 ,
\end{equation}
which is never negative for $0<\eta_0<\pi f_1$ and is only zero at  $\eta_0^{m} \equiv 2 \pi m q f_1$, with $m=\{0,1,2\dots\}$. Close to these points the sign of the derivative will come from terms with higher powers of $q$. The one multiplying $q^{N+2}$ is:
\begin{equation}\label{e:derVsines}
\sin{\left(\eta_0^m\over f_2\right)} \approx {\eta_0 \over q f_1} - 2 \pi m.
\end{equation}

\begin{figure}[h]
\includegraphics[width= \columnwidth]{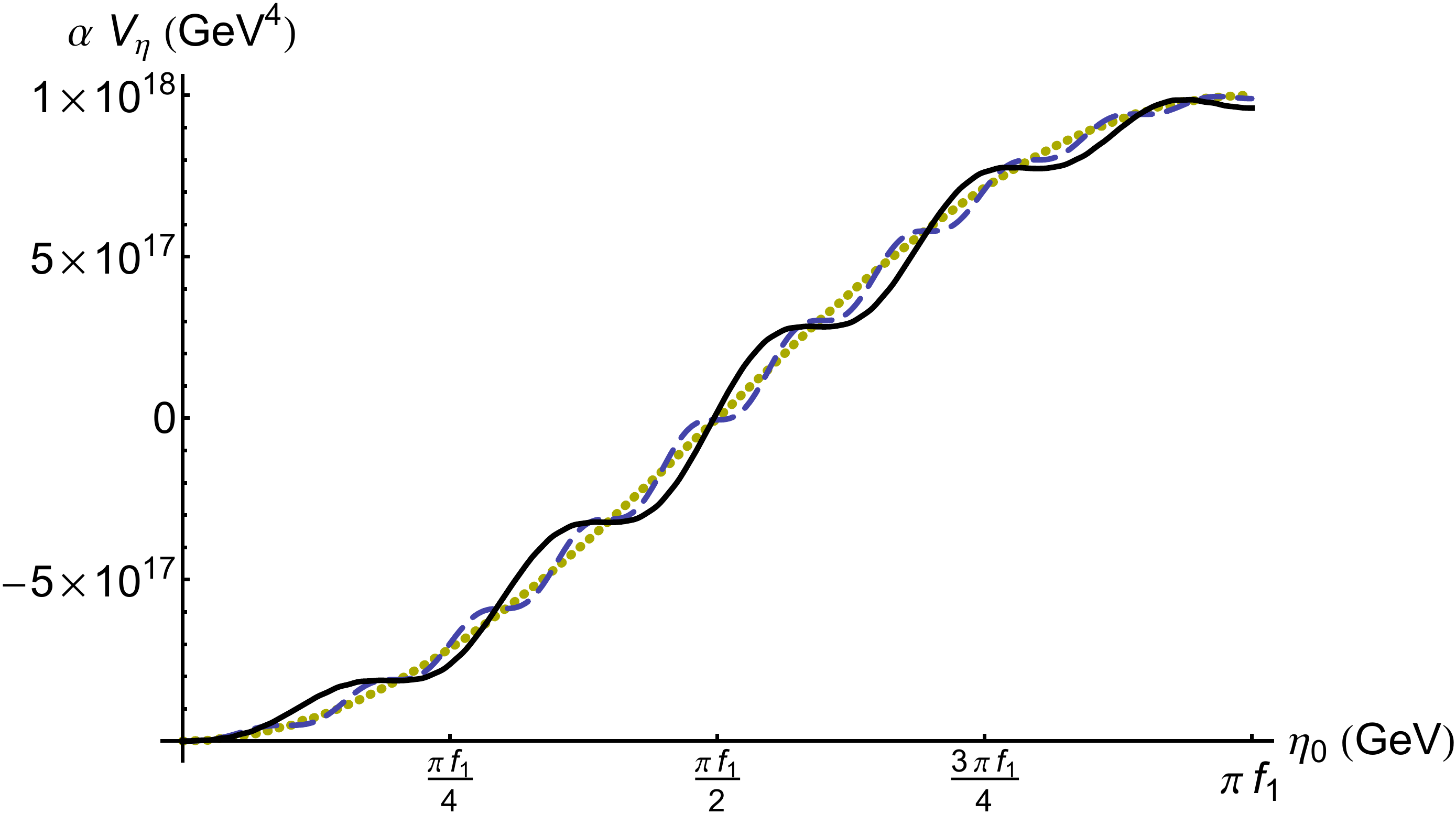}
\caption{\label{fig:pot_vs_q} %\footnotesize
Potential $V_{\eta}(\eta_0)$ for different values of $q$ and $N=3$. A  factor
$\alpha$ was introduced to allow easy comparison between the curves. The black, blue (dashed) and yellow (dotted) curves have respectively $(q=0.1,  \alpha = 1)$, $(q=0.05, \alpha = 10^3)$ and $(q=0.01, \alpha=10^{10})$. Note that these values of $q$ are much larger than the realistic ones, in order to exacerbate the features in the potential.}
\end{figure}

This sine will push the derivative to negative values near  $\eta_0^{m}$, generating shallow minima (similar arguments apply to the next terms in the $q$-expansion). The derivative only remains negative while the term in \Eq{e:derV1} is smaller than the $\mathcal{O}(q^{N+2})$ term, so these minima become less and less important as $q$ gets smaller. In fact, the height of the barrier
between two adjacent minima decreases as $q^4$, the width decreases as $q^{2-N} \mathcal{C}_N$ and we expect the field to be able to proceed rolling down for the typical values of $q$ considered below. The shape of the potential with decreasing $q$ can be seen in Figure~\ref{fig:pot_vs_q}. One can see that, despite the use of quite large values of $q$ and a scaling factor $\alpha$ to exacerbate the features of the potential, the slope quickly gets  smooth.

We  include the Higgs by multiplying the Lagrangian by  $1+ |H|^2/\Lambda^2$, where $\Lambda \approx 4 \pi f$.  Adding in the Higgs potential and kinetic term, the full Lagrangian is now:
\begin{equation}
\mathcal{L}_{\eta,H}=\left(1+ \frac{|H|^2}{\Lambda^2}\right)\mathcal{L}_\eta +|D_\mu H|^2+\frac{\Lambda^2}{2} |H|^2 -\frac{\lambda_H}{4} |H|^4.
\end{equation}

Once the Higgs is set to its vev, $\langle h \rangle = v>0$,
the slope equation is the same as \Eq{e:derV}, multiplied by $(1+v^2/(2 \Lambda^2))$. The field $\phi$ should stop rolling when this expression is approximately zero. However, this clearly has no solutions apart from the trivial one $v^2=- 2 \Lambda^2$, which is undesirable.

With the current Lagrangian, having $v \ll f$ is untenable. In order to fix this,
 we add the following breaking term at the last site, which can be generated by the UV completion shown in \Eq{e:uvhiggs} or would be equivalent, in the continuum  limit, to a deformation of the metric in the infrared (IR), as in Appendix \ref{app:pngb}:

\begin{equation} \label{e:eps}
\mathcal{L}_{\eta,H} \rightarrow \mathcal{L}_{\eta,H} +  \epsilon \frac{\Lambda_c}{16 \pi} \mathrm{Tr}[\Phi_N+\Phi_N^\dag] |H|^2
\end{equation}
where $\epsilon$ is a small parameter, and $\Lambda_c$ is a new scale, which we  could assume is generated at lower energies to avoid spoiling the results of the previous section (see \Eq{e:relaxion}). However, as pointed out in \cite{Espinosa:2015eda}, a small $\Lambda_c$ scale leads to a coincidence problem (i.e., $\Lambda \gg \Lambda_c \sim \,$TeV) for the model. We will then take $\Lambda_c \approx \Lambda\approx 4\pi f$ and discuss below how to avoid the problems generated by this choice.

Once this operator is added, the relaxion potential acquires the term
\begin{equation} \label{e:epsl}
\epsilon f^2 |H|^2 \cos{\frac{\eta_0}{f_N}},
\end{equation}
giving the relaxion a small mass. By closing the loop of $H$,  $\epsilon f^4 \cos (\eta_0/f_N)$ is generated, which can spoil the relaxation mechanism. One possible solution is to adopt the double scanner mechanism of \cite{Espinosa:2015eda}, that is, we may add a scalar singlet  to control the amplitude of the additional term.
As emphasized in \cite{Espinosa:2015eda}, the new field needs an even larger field excursion than the relaxion. This can be easily accommodated in our framework by replicating this scalar on the $N$-sites, provided we choose a smaller value of the $q$ parameter for this scalar. A non-trivial issue that must be addressed in a complete model is the fact that the UV completion should not couple the new scalar to the Higgs at tree level, or else one risks spoiling the relaxation \cite{Espinosa:2015eda}. We hope it is feasible to overcome this difficulty with clever model building, however, the details of this construction and the continuum limit thereof are beyond the scope of our paper and left for future work. For a supersymmetric version of a two-field relaxion model, see \cite{Evans:2016htp}.

With the inclusion of (\ref{e:epsl}), the new slope equation is given by:
\begin{align}\label{e:slope2}
&\frac{\partial V_{\eta,H}}{\partial \eta_0}=\frac{f^3 q^{N+1}}{\mathcal{C}_N} \left\{ \left(1+\frac{v^2}{2 \Lambda^2}\right)\sin{\left(\frac{\eta_0}{f_1}\right)}\bigg[1-\cos{\left(\frac{\eta_0}{f_2}\right)}\right.\nonumber\\
&\left.+\mathcal{O}(q)\bigg]-\epsilon \frac{ v^2}{2 f^2 q^{N+1}} \sin{\left(\frac{\eta_0}{f_N}\right)}\right\}+\cdots.
\end{align}

This slope should be zero when $v \approx 246$ GeV. %the vev of $H$ is of the order of the electroweak scale.
Solving for this yields
\begin{equation}\label{e:vev}
v^2 \sim \frac{f^2}{\epsilon} q^{N+1}\,.
\end{equation}

For  $q^{N+1}<\epsilon <1$, a natural electroweak scale is obtainable and  $q^{N+1}$ should be identified with the relaxion coupling $g$ of \cite{Espinosa:2015eda}, as in \Eq{e:Vrelaxion}.

The cutoff for our model can be estimated along the lines of \cite{Espinosa:2015eda} by considering additional constraints besides \Eq{e:vev}. The main bounds come from requiring that $\vec{\eta}_0$ does not drive inflation, i.e. $\Lambda^2 \lesssim H_I M_{\text{Pl}} $, where $H_I$ is the inflation scale and $M_{\text{Pl}}$ is the reduced Planck scale, and that quantum fluctuations of $\vec{\eta}_0$ are less important than its classical rolling. This yields the condition that $H_I^3 \lesssim q^{N+1} f^3$. Finally, suppressing higher order terms like $\epsilon^2 f^4 \cos(\eta_0/f)^2$ requires $\epsilon \lesssim v^2/f^2\sim 10^{-12}$, for $f=10^{8}~$GeV \cite{Espinosa:2015eda}. Combining these with \Eq{e:vev}, we obtain:
\begin{equation}
  \frac{\Lambda^6}{f^3 M_{\text{Pl}}^3}\lesssim q^{N+1} \lesssim\frac{v^4}{f^4}. %
\end{equation}
From this, we find the upper bound of $f \lesssim 10^8$ GeV and also that  $q \lesssim 10^{-23/(N+1)}$.

Finally, using all these constraints, we find that for $q\approx 10^{-24/(N+1)}$  and $\epsilon \approx 10^{-12}$,
we obtain $v \sim 10^{-6} f$ which is of the order of the electroweak scale for $f \approx 10^{8}$~GeV. Note that for these parameter choices, \Eq{e:vev} does not depend on $N$. Of course, having a large value for $N$ allows for a larger value of $q$.

%%%%%%%%%%%%%%%%%%%%%%%%%%%%%%%%%%%%%%%%%%%%%%%%%%%%%%%%%%%
%%%%%%%%%%%%%%%%%%%%
\section{\label{s:conclusion} Discussion}

We have constructed a simple $2N$-site model capable of addressing two problematic points of the relaxation mechanism, namely the necessity for (i) large field excursions and (ii) a linear term that explicit breaks the axion shift symmetry. Our model generates a potential  composed of many oscillatory terms with very different periods (see \Eq{e:Leta}), the term with the larger period plays the role of the linear term in \Eq{e:Vrelaxion}. From $N$ fields acquiring expectation values of order $f$, an effective scale $f_1 = \mathcal{C}_N f/q^{N-1} \gg f$ (see \Eq{e:fjota}) is generated and the pNGBs have a compact field space of $2 \pi f_1$, which allows for large field excursions.

The present model has some distinctive features when compared with previous many-field models that also address the points above~\cite{Kaplan:2015fuy,Choi:2015fiu}:
\begin{itemize}
\item
The $N$ fields are bi-fundamentals of $2N$ non-abelian $SU(2)$ groups and the formalism employed can be trivially generalized to any non-abelian group. This allows for a controlled UV behavior and opens up many possibilities of model building in particle physics and inflation.

\item
The model has a well defined continuum limit $N \rightarrow \infty$, $q \rightarrow 1$, with $q^{N+1}$ kept fixed, and the mass matrix for the pNGBs in \Eq{m:MNG2} is exactly the one obtained  from a pNGB Wilson line in the deconstruction of AdS$_5$ \cite{Burdman:2014ixa, tese:2013} (see Appendix \ref{app:pngb}). Even the desired relation between $v$ and $f$ (in \Eq{e:vev}) is maintained in the continuum limit, as $f^2 q^{N+1}\rightarrow M/g_5^2~e^{-k L}$, where $L$ is the size of the extra dimension, $k$ is the curvature, $g_5$ is the $5d$ gauge coupling, and $M$ is the cutoff of the UV theory~(see Appendix \ref{app:UV}). In addition, we find that (up to suppressed terms) in the continuum limit (see \Eq{e:fjota}), $f_1= \mathcal{C}_N f q^{1-N} \rightarrow M/(g_5 \sqrt{2k})e^{ k L}$ and $f_N= \mathcal{C}_N f \rightarrow M/(g_5 \sqrt{2k})$, that is $f_1/f_N\rightarrow e^{kL}$, i.e. they are related by the AdS$_5$ warp factor. These expressions are in agreement with those obtained by \cite{Falkowski:2006vi} in AdS{$_5$}.
\end{itemize}

While the potential of \Eq{e:Leta} has shallow minima that do not affect the slow roll of the relaxion, adding the Higgs requires the introduction of a new term that generates large barriers for $\langle H\rangle \neq 0$. The extra breaking is proportional to a new spurion $\epsilon$ and ultimately controls the magnitude of the Higgs vev via \Eq{e:vev}. In the continuum limit, this should correspond to an IR deformation of the extra dimensional metric. This operator may also spoil the relaxation mechanism  via higher order corrections, but we expect these can be amended by adopting the double scanner scenario of \cite{Espinosa:2015eda}.

In the viable region of parameter space, we find that the cutoff of the model can be pushed up to $\Lambda\approx4\pi f\sim 10^9~$GeV.

The breaking term of \Eq{e:eps} is not unique, and it may be possible to avoid introducing it by considering different terms in \Eq{e:Lphi} that automatically generate the large barriers needed to stop the rolling of $\vec{\eta}_0$. Alternatively, one might be able to achieve the same result through changing the parametrization of the $g_j$ couplings in the Lagrangian in order to mimic a metric that is slightly deformed from AdS{$_5$}.

 It will also be interesting to investigate the continuum limit of this model (i.e.  a warped extra dimension), which is a possible direction to achieve an UV completion that is compatible with the WGC \cite{delaFuente:2014aca}.

 Additionally, the framework established here could find application in model building of the inflation sector, which also requires large field excursions, for instance, in models with observable primordial gravitational waves \cite{Lyth:1996im}.

%%%%%%%%%%%%%%%%%%%%%%%%%%%%%%%%%%%%%%%%%%%%%%%%%%%%%%%%%%%
%%%%%%%%%%%%%%%%%%%%
\section*{Acknowledgments}

This material is based upon work supported by the S\~{a}o Paulo Research Foundation
(FAPESP) under grants 2013/01907-0, 2012/21436-9, 2012/21627-9 and 2015/15361-4.
The authors would like to thank Gustavo Burdman, Eduardo Pont\'{o}n,  Pedro Schwaller, and specially Christophe Grojean and G\'{e}raldine Servant  for useful discussions.

%%%%%%%%%%%%%%%%%%%%%%%%%%%%%%%%%%%%%%%%%%%%%%%%%%%%%%%%%%%
%%%%%%%%%%%%%%%%%%%%
\appendix

%%%%%%%%%%%%%%%%%%%%%%%%%%%%%%%%%%%%%%%%%%%%%%%%%%%%%%%%%%%%
%%%%%%%%%%%%%%%%%%%%%
\section{\label{app:UV} Fermionic UV Model}

%The Lagrangian for the link fields used in the main text :
%
%\begin{align}\label{e:Lphi}
%\mathcal{L}_{\Phi}&=\sum_{j=1}^N \left\{ \mathrm{Tr}\left[(\partial_\mu \Phi_j)^\dag \partial^\mu \Phi_j\right]+\right.\nonumber\\
%&\left.+\,\frac{f^3}{2}(2-\delta_{j,1}-\delta_{j,N})g_j^2 \mathrm{Tr}\left[\Phi_j+\Phi_j^\dag\right] \right\}
%\nonumber\\ &- \frac{f^2}{2}\sum_{j=1}^{N-1} g_j g_{j+1} \mathrm{Tr}\left[(\Phi_j- \Phi_{j}^{\dagger})(\Phi_{j+1}- \Phi_{j+1}^{\dagger} ) \right],
%\end{align}
%
The Lagrangian of \Eq{e:Lphi} can be generated by a simple UV model, obtained by $2N$ multiplets of Dirac fermions, transforming as $SU(2)$ doublets, at a high energy scale, with the following lagrangian:
\begin{align}\label{e:Luv}
\mathcal{L}_{UV}&=\sum_{j=1}^{N} \left\{\bar{\psi}_j \slashed{p} \psi_j+\bar{\chi}_j \slashed{p} \chi_j\right\}\nonumber \\
&+\sum_{j=1}^{N-1}\left\{ \bar{\psi}_{Lj} \left[\lambda_j \phi_j+\lambda_{j+1} \phi_{j+1}-\lambda^\prime_j f \right]\psi_{Rj} \right.\nonumber \\
&\left.+\bar{\chi}_{Lj} \left[\tilde{\lambda}_j \phi_j-\tilde{\lambda}_{j+1} \phi_{j+1}^\dag-\tilde{\lambda}^\prime_j f \right]\chi_{Rj}+\mathrm{h.c.}\right\},
\end{align}
where $L,~R$ denote chirality projections and the couplings $\lambda_j,~\lambda_j^\prime,~\tilde{\lambda}_j, ~\tilde{\lambda}^\prime_j$ are assumed small. Upon integrating out these fermions and matching the couplings, one obtains the Lagrangian of \Eq{e:Lphi}, plus terms suppressed by higher orders of the couplings.

The additional term introduced in  \Eq{e:eps}  can be similarly generated by
\begin{equation}\label{e:uvhiggs}
\mathcal{L}_{UV}^\prime= \xi^\dag \slashed{p} \xi +\zeta \slashed{p} \zeta^\dag+  \xi(\epsilon \phi_N -m )\zeta +\mathrm{h.c.},
\end{equation}
where $\xi,~\zeta$ are a set of chiral fermions located at the last site.
The Higgs may then be added trivially by multiplying the entire Lagrangian by the EW singlet $1+H H^\dag/\Lambda^2$.

%%%%%%%%%%%%%%%%%%%%%%%%%%%%%%%%%%%%%%%%%%%%%%%%%%%%%%%%%%%%
%%%%%%%%%%%%%%%%%%%%%
\section{\label{app:pngb} pNGB Wilson line in deconstructed $\text{AdS}_5$ }

Consider the action for the gauge field of a group $\mathcal{G}$ in a slice of $\text{AdS}_5$ in proper coordinates \cite{Gherghetta:2000qt}, $ds^2=e^{-2 k y}\eta_{\mu \nu} dx^\mu dx^\nu -dy^2$:

\begin{align}\label{S5continua}
S_5^A &= \int d^4 x \int^{\pi R}_0 dy \sqrt{-g}\left\{- \frac{1}{2 g_5^2} \mathrm{Tr} \left[F_{MN}^2\right]\right\} \nonumber \\
&= \int d^4 x \int^{\pi R}_0 dy  \left\{-\frac{1}{2g_5^2} ~\mathrm{Tr}\left[ F_{\mu \nu}F^{\mu \nu}\right] \right.\nonumber\\
&\left.+\frac{1}{g_5^2}e^{-2ky}~\mathrm{Tr}\left[(\partial_5 A_\mu - \partial_\mu A_5)^2\right] \right\}\mathrm{.}
 \end{align}

We discretize the extra dimension by substituting
\begin{eqnarray}
&\int^{\pi R}_0 dy~& \rightarrow \sum_{j=0}^N a \nonumber \mathrm{,} \\ \nonumber \\
&\partial_5 A_\mu& \rightarrow \frac{A_{\mu,j}-
 A_{\mu,j-1}}{a}\mathrm{,}
 \end{eqnarray}
 where $a$ is the lattice spacing (inverse cutoff). We obtain:
%{\small
\begin{align}\label{f:S51}
S_5^A = \frac{a}{g_5^2}\int d^4 x \left\{ -\frac{1}{2}
\sum_{j=0}^N \mathrm{Tr}\left[F_{\mu\nu, j} F_j^{\mu \nu}\right]\right.\nonumber\\
\left. + \sum_{j=1}^N \frac{e^{-2k a j}}{a^2}   \mathrm{Tr}\left[\left(A_{\mu,j}-
 A_{\mu,j-1} -a \partial_{\mu}A_{5,j}\right)^2\right]\right\}
\mathrm{.} \end{align}%}

Consider now a theory of $N+1$ gauged non-linear sigma model fields, $U_j$. The scalar fields act like linking fields in a lattice, transforming under adjacent gauge groups (assumed to be all equal to $\mathcal{G}$) as $U_j \rightarrow L_j U_j R_j^\dag$, where $L_j, ~R_j$ are the gauge symmetries on sites $j,~j+1$, respectively. The $U_j$ spontaneously break $L_j \times R_j \rightarrow V_j$ at an scale $f_j$, yielding $N+1$ multiplets of NGB fields, $\pi_j$. We can match the discretized action above to this gauged non-linear sigma model action, by expanding it at the quadratic level in the Nambu-Goldstone fields:
\begin{align}\label{f:S42}
S_4^A = \frac{1}{g^2}\int d^4 x \left\{ - \frac{1}{2 }\sum_{j=0}^N \mathrm{Tr}\left[ F_{\mu\nu,j
}~F^{\mu\nu}_{j}\right] + \right.\nonumber\\
\left.\sum_{j=1}^N f^2 g^2 q^{2j}\mathrm{Tr}\left[\left(A_{\mu,j}-
A_{\mu,j-1}-\partial_{\mu}\frac{\pi_j}{f_j} \right)^2  \right]\right\}
\mathrm{,} \end{align}
where $\pi_j$ is a Goldstone mode transforming in the adjoint of the vector symmetry $V_j$, and we take $f_j \equiv f q^j$, by making the identifications \cite{ArkaniHamed:2001ca,Hill:2000mu,Bai:2009ij,deBlas:2006fz}:

\begin{align}
\label{e:dicionario}
&\frac{g_5^2}{a} \leftrightarrow g^2\mathrm{,} \nonumber \\
&f \leftrightarrow \frac{1}{\sqrt{a} g_5}=\frac{1}{a g}\mathrm{,} \nonumber \\
&q \leftrightarrow e^{-ka},
\end{align}
we see the Goldstone mode is identified with the scalar component of the gauge field. Or, equivalently, the non-linear linking field $U_j = e^{i \pi_{j}/ f_j}$ is identified with the Wilson line $\exp\left[{i\int_{aj}^{a(j+1)} dy \,A_5 e^{-2 k y}}\right]$.

Now, consider the breaking $\mathcal{G}\rightarrow\mathcal{H}$ by boundary conditions in theory space, that is, we assume that the first and last sites, the symmetry group is reduced to $\mathcal{H}$. Alternatively, we can implement this breaking by localized scalar fields, then take their vev to infinity, decoupling the massive gauge modes.

Denoting the broken generators by hatted indexes, it is straightforward to see that we can remove the mixing between Goldstone modes and gauge fields by adding the gauge fixing term:
\begin{eqnarray}\label{rxih}%
\mathcal{L}_G&=& -\! \sum_{j=1}^{N-1} \!\frac{1}{2\xi} \left[\partial_\mu A^{\mu,\hat{a}}_{j} \!+ \xi  \left( f_j \pi_j^{\hat{a}} -f_{j+1} \pi_{j+1}^{\hat{a}} \right)  \right]^2\!\!\!.
\end{eqnarray}
\noindent One may then verify that the mass matrix obtained for the NGB fields parametrizing $\mathcal{G}/\mathcal{H}$ is given by: \cite{Burdman:2014ixa, tese:2013}
% %
% \scriptsize
\begin{equation}\label{mpih}
M_\pi^2 =f^2 \xi \left( \begin{array}{cccccc}
~~q^2     & -q^3      & ~~0        &\cdots   &~~0                 & ~~0 \\
-q^3      & 2q^4      & -q^5        &\cdots   &~~0                 & ~~0 \\
~~0       & -q^5      & 2 q^6    &\cdots   &~~0                 & ~~0 \\
\vdots    & \vdots    & \vdots   &\ddots   &\vdots              & \vdots \\
~~0       & ~~0       & ~~0         &\cdots   &2q^{2(N-1)}  & -q^{2N-1}   \\
~~0       & ~~0       & ~~0        &\cdots   &-q^{2N-1}     &q^{2N}
\end{array} \right)
\mathrm{.} \end{equation}
%\normalsize
reproducing the mass matrix obtained in \Eq{m:MNG2}. Note that while the massive modes have gauge dependent masses, the zero mode is physical.

\bibliography{NRelBib}

\end{document}